# Sulfur-rich Spirofluorene-Bridged N Heterotriangulene Redox-Active Polymers

Angelina Jocic[a], Tom Wickenhäuser[b], Sebastian Lindenthal[c], Alexander Welle[d], Vanessa Trouillet[d],Ronald Curticean[e], Irene Wacker[e], Jana Zaumseil[c], Rasmus R. Schröder[e], Rüdiger Klingeler[b], Milan Kivala[a,1]

[a]Institute of Organic Chemistry, Heidelberg University, Im Neuenheimer Feld 270, 69120 Heidelberg (Germany)

[b]Kirchhoff-Institute for Physics, Heidelberg University, Im Neuenheimer Feld 227, 69120 Heidelberg (Germany)

[c]Institute for Physical Chemistry, Heidelberg University, Im Neuenheimer Feld 253, 69120 Heidelberg (Germany)

[d]Institute for Functional Interfaces and Karlsruhe Nano Micro Facility, Karlsruhe Institute of Technology,Hermann-von-Helmholtz-Platz 1, 76344 Eggenstein-Leopoldshafen (Germany).

[e]BioQuant, Heidelberg University, Im Neuenheimer Feld 267, 69120 Heidelberg (Germany)

**Abstract:**

Sulfur-rich spirofluorene-bridged *N*-heterotriangulene (FTN) polymers featuring covalently linked oligosulfide units and a terthiophene-based analogue were synthesized via nucleophilic aromatic substitution and Stille cross-coupling polymerization. The resulting materials are amorphous, insoluble solids with high thermal stability and sulfur contents up to 25 wt%. Structural and compositional analyses by combustion analysis, ToF-SIMS, FT-IR, XPS, and solid-state NMR confirm the efficient incorporation of short oligosulfide to disulfide linkages and well-defined terthiophene units in the respective polymers. Electrochemical characterization in lithium half-cells reveals a reversible, high-voltage oxidation of the FTN unit at 3.8–4.0 V (vs. $Li/Li^+$), accompanied by low-voltage sulfur- or terthiophene-based redox processes between 1.5–2.5 V (vs. $Li/Li^+$). Sulfur incorporation markedly increases the theoretical and initial discharge capacities (up to 129 mA h $g^{-1}$), while the sulfide conversion processes exhibit rapid fading and poor reversibility due to sulfide dissolution. In contrast, the terthiophene-linked polymer shows only transient low-voltage activity while maintaining high Coulombic efficiencies (≈99.7%) governed by the persistent FTN backbone redox event. Our results highlight how different redox-active linkers influence the electrochemical behavior of FTN-based polymers and provide insights into the design of functional organic cathode materials featuring multi-redox processes.

[1] milan.kivala@oci.uni-heidelberg.de

## 1. Introduction

Sulfur is one of the most abundant and versatile elements in the crust of the Earth and exhibits remarkable chemical diversity owing to its ability to adopt multiple oxidation states and to form extended S–S chains.[1–4] In organic electronic materials, the incorporation of sulfur enhances polarizability and intermolecular interactions, which are crucial for molecular packing and charge transport.[1,5,6] In thiophene, the sulfur atom stabilizes the aromatic π-system and enables versatile chemical modifications, thereby establishing thiophene as a key motif in modern organic semiconductor materials.[7–10] Beyond optoelectronics, sulfur also exhibits soft-ligand properties with rich redox chemistry, enabling broad application in metal binding,[11–13] catalysis,[14,15] and energy storage systems such as lithium-sulfur batteries.[16–18] Elemental sulfur (as $S_8$) undergoes a multielectron conversion reaction with lithium to form $Li_2S$, resulting in a high theoretical specific capacity of ~1672 mAh $g^{-1}$ (Figure 1A).[16,19,20] In ether-based electrolytes, discharge proceeds via the stepwise formation of soluble lithium polysulfides ($Li_2S_x$), giving rise to characteristic two-step discharge plateaus.[21,22] Despite these attractive cathode properties, elemental sulfur is limited due to dissolution of small oligosulfide fragments into the electrolyte, which induces redox imbalance at the lithium anode and leads to rapid capacity fading with low Coulombic efficiency (CE).[16,23] To address the intrinsic limitations of elemental sulfur, various strategies have been developed to stabilize sulfur and its reactive sulfur chains.[24–27] One effective approach is covalent fixing, in which sulfur atoms or chains are chemically bound to organic frameworks.[28–30] Such covalently fixed structures provide a homogeneous distribution of electroactive sulfur sites, suppress oligosulfide migration, and enhance long-term cycling stability.[27,31,32] Several synthetic approaches have been developed to incorporate sulfur covalently into polymeric frameworks.[33–36] A widely used method is inverse vulcanization, in which molten sulfur reacts with unsaturated organic comonomers to form sulfur-rich networks.[37–39] The resulting polymers contain up to 90 wt% sulfur and typically deliver reversible capacities of 600–1000 mAh $g^{-1}$ with excellent cycling stability.[36,40,41] Beyond these thermally induced reactions, solution-based nucleophilic aromatic substitution ($S_NAr$) has emerged as an alternative route to sulfur-rich polymers under comparatively mild conditions.[42–45] Nevertheless, the number of reported polymer systems prepared through $S_NAr$ remains limited and the scope of accessible structures is still developing. An alternative strategy to mitigate the drawbacks of elemental sulfur involves embedding sulfur directly into the π-conjugated backbone, thereby creating a redox-active heteroaromatic system that can reversibly store charge without forming soluble polysulfides.[46–50] Although oligothiophenes such as terthiophene (**TTP**) are among the most extensively studied structural motifs in organic electronics,[51] they remain surprisingly underexplored as molecular redox units.[52–54] The sulfur-embedded, unsubstituted aromatic framework in **TTP** enables one-electron storage at comparatively negative potentials, forming a radical anion of limited stability (Figure 1A).[55,56] However, chemical modification of the **TTP** core can substantially tune its redox properties, enabling substituted derivatives to access two-electron redox processes.[52]

In 2024, our group developed a redox-active polymer based on spirofluorene-bridged *N*-heterotriangulenes (FTNs), planarized triphenylamine-derivatives featuring rigid C($sp^3$)-spiro-bridges that effectively stabilize the nitrogen-centered radical cation (Figure 1B).[57,58] The corresponding FTN-based polymers (**FTN-Pol** and **FTN-H-Pol**), obtained via Yamamoto polymerization, are amorphous and microporous with surface areas around 690 $m^2$ $g^{-1}$ and excellent thermal stability. Electrochemical measurements revealed a reversible nitrogen-centered redox event at 3.8–3.9 V (vs. Li/$Li^+$), delivering stable specific capacities of up to 26 mAh $g^{-1}$ and outstanding cycling stability over 400 cycles. These results established spiro-bridged *N*-HTA frameworks as robust π-conjugated platforms for organic cathode materials.[58]

Motivated by these findings, we aimed herein to expand the series of FTN-based polymers as redox-active materials toward sulfur-containing and terthiophene-linked derivatives to explore the influence of additional redox-active linkers on the overall redox behavior and charge-storage properties. Specifically, two sulfur-containing FTN polymers featuring oligosulfide linkages and a terthiophene-linked analogue were synthesized and isolated as amorphous insoluble solids which were subjected to comprehensive structural and spectroscopic analyses. Their performance as cathode materials in lithium-ion half-cells was evaluated.

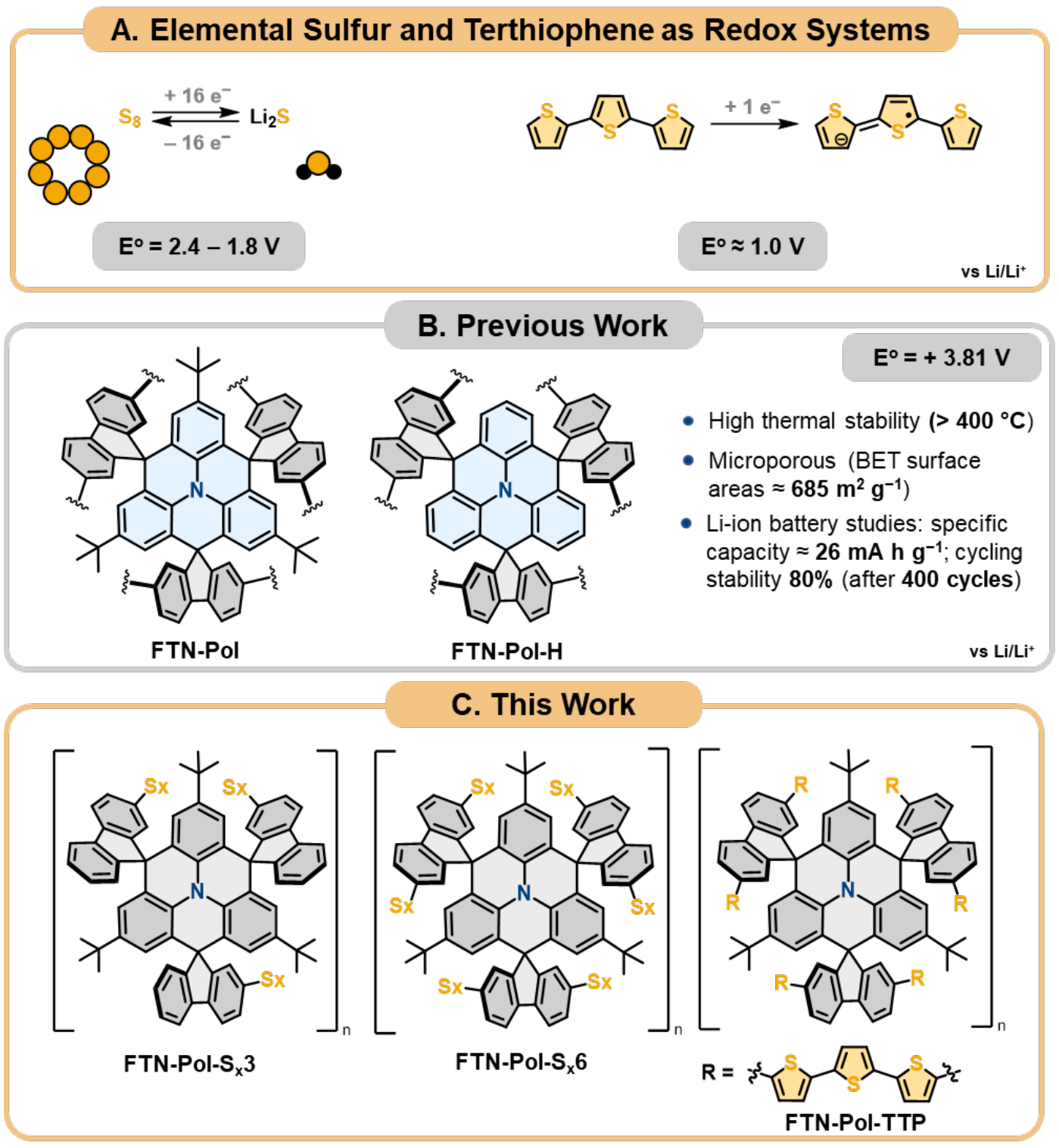

**Figure 1:** (A) Redox behavior of elemental sulfur and terthiophene (TTP). (B) Previously reported FTN polymers featuring high-voltage, reversible nitrogen-centered redox process. (C) This work: **FTN-Pol** derivatives with sulfur-based (**FTN-Pol-S$_x$3**, **FTN-Pol-S$_x$6**) or **TTP** linkers (**FTN-Pol-TTP**), enabling additional redox activity while preserving the **FTN-Pol** high-voltage core.

## 2. Results and Discussion

### *2.1. Synthesis of the Brominated Monomers*

The synthesis of the brominated monomers **1a** and **1b** is shown in Figure 2A. While the 6-fold brominated compound **1b** had already been established by us previously, the 3-fold brominated compound **1a** was newly synthesized by a modified protocol.[58,59] Triphenylamine **2** was subjected to lithium/bromine exchange with $^n$BuLi at −78 °C and subsequent addition of 2-bromofluorenone or 2,7-dibromofluorenone afforded the respective triols **3a** or **3b** in yields between 20−25%. Threefold cyclization with trifluoromethanesulfonic acid at 0 °C afforded the spiro-compounds **1a** and **1b** in a yield of 80 and 90 %, respectively. Both brominated monomers are colorless, crystalline solids, with high thermal stability and, importantly, they are accessible in gram quantities (for more details, see the Supporting Information).

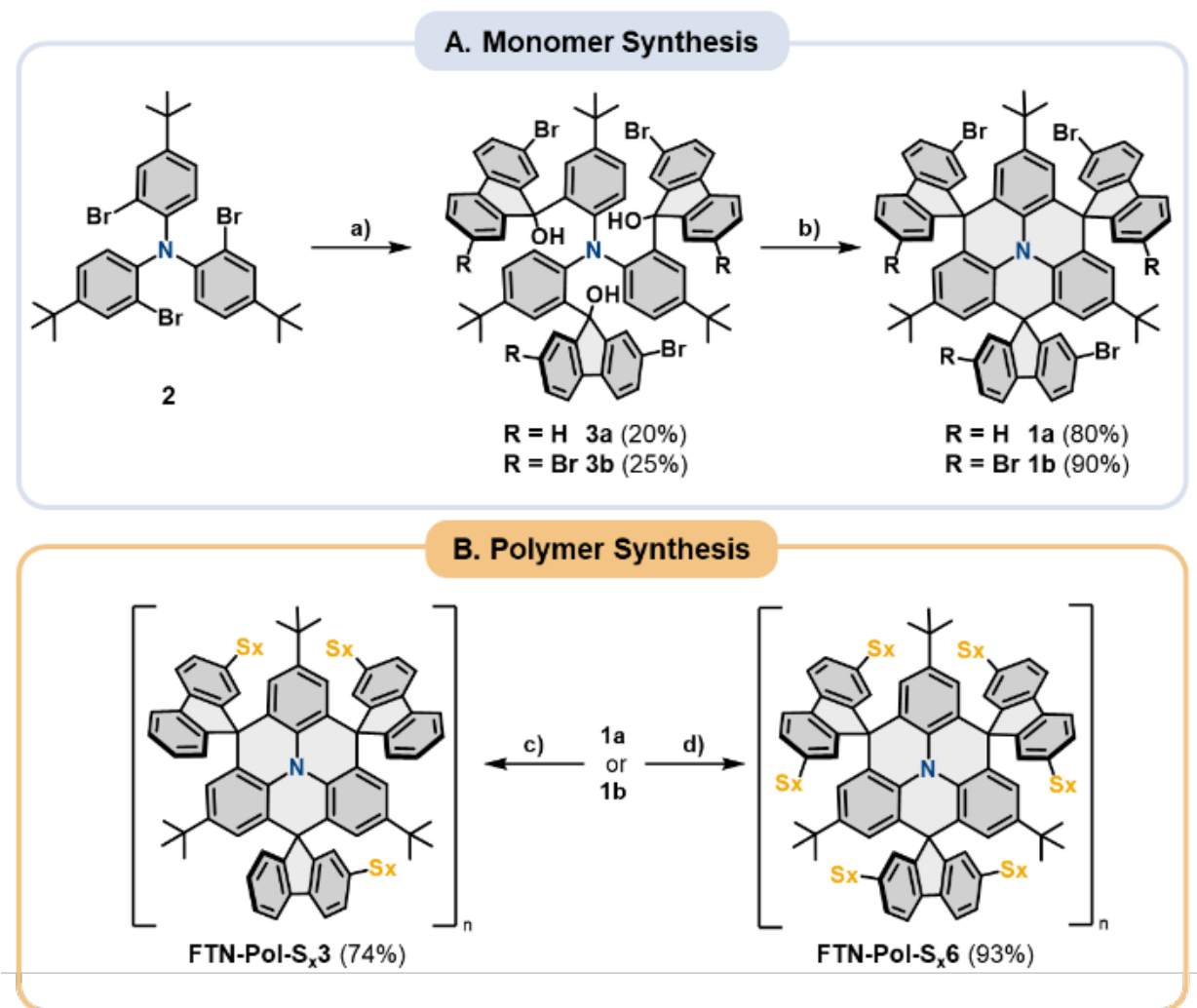

**Figure 2:** Synthetic route to the brominated spirocyclic monomers **1a** and **1b** (A) and the corresponding sulfur-rich polymers **FTN-Pol-S$_x$3** and **FTN-Pol-S$_x$6** (B). Reagents and conditions: a) 1) $^n$BuLi, THF, −78 °C, 20 min, $N_2$; 2) 2-bromofluorenone (for **3a**) or 2,7-dibromofluorenone (for **3b**), THF, −78 °C to room temperature, 12 h, $N_2$; b) $CF_3SO_3H$, $CHCl_3$ (for **1a**) or $CF_3SO_3H$ (for **1b**), 0 °C, 5 min; c) and d) 1) $Na_2S$, $S_8$, NMP, 180 °C, 20 min, $N_2$; 2) NMP, 180 °C, 48 h, $N_2$. THF = tetrahydrofuran.

### *2.2. Synthesis and Characterization of the Sulfur-linked Polymers **FTN-Pol-S$_x$3** and **FTN-Pol-S$_x$6***

The sulfur-rich polymers **FTN-Pol-S$_x$3** and **FTN-Pol-S$_x$6** were synthesized via $S_N$Ar of the brominated spiro-compounds **1a** and **1b** with *in situ* generated sodium oligosulfide species (Figure 2B, for more details see the Supporting Information).[60,61] The reactive sodium oligosulfide solution was obtained by heating sodium hydrosulfide hydrate with elemental sulfur in *N*-methyl-2-pyrrolidone (NMP) at 180 °C and was directly treated with the respective brominated monomer **1a** or **1b**. Purification by Soxhlet extraction ($H_2O$, methanol, THF, and benzene) and subsequent drying under reduced pressure at 120 °C afforded **FTN-Pol-S$_x$3** and **FTN-Pol-S$_x$6** as black solids in good to excellent yields of 74 and 93%, respectively (see Figure 2A). Both polymers are insoluble in water and common organic solvents.

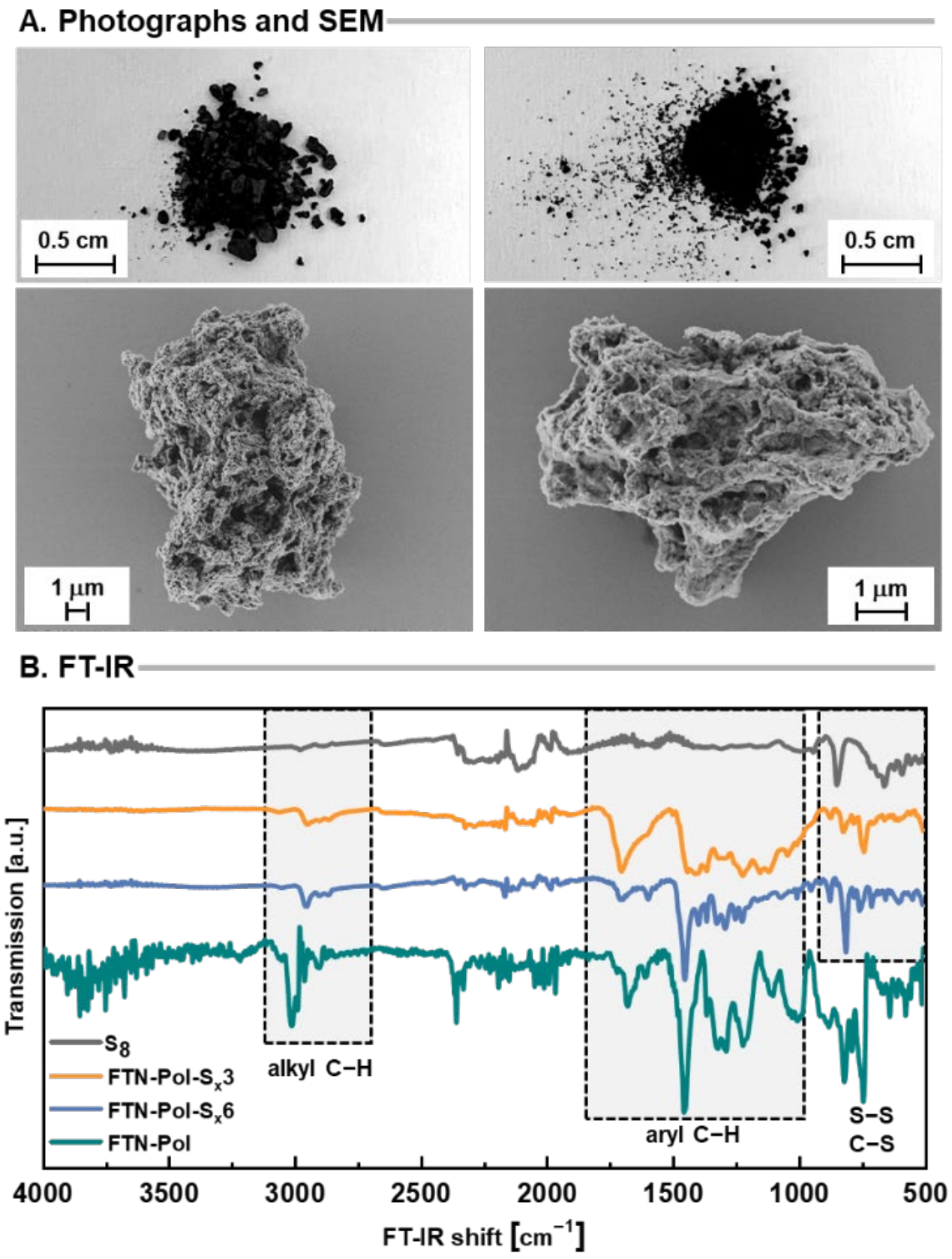

**Figure 3:** Morphological and structural characterization of **FTN-Pol-$S_x$3** (left) and **FTN-Pol-$S_x$6** (right). (A) Photographs (top) and SEM images (bottom) of the powders; (B) FT-IR spectra of the polymers and the model compounds **FTN-Pol** and elemental sulfur ($S_8$).

Photographs of the purified polymers reveal their dark color and solid amorphous morphology after extensive purification and drying (Figure 3A, top). To elucidate the elemental composition, purity, and polymerization degree of **FTN-Pol-$S_x$3** and **FTN-Pol-$S_x$6**, elemental combustion analysis, time-of-flight secondary ion mass spectrometry (ToF-SIMS), powder X-ray diffractometry (PXRD), and X-ray photoelectron spectroscopy (XPS) were performed. Further structural characterization was achieved by Fourier-transform infrared (FT-IR), Raman, and cross-polarization magic angle spinning (CP-MAS) $^{13}$C NMR spectroscopy. In addition, the thermal stability and morphology were investigated via thermogravimetric analysis (TGA), scanning electron microscopy (SEM), and nitrogen sorption measurements, respectively.

Elemental analysis revealed sulfur contents of 24.7 wt% for **FTN-Pol-$S_{x3}$** and 23.6 wt% for **FTN-Pol-$S_{x6}$**, confirming the efficient sulfur incorporation. Based on the molecular structure of the repeating FTN unit, these values correspond to approximately one to two sulfur atoms per peripheral substituent, i.e. six to twelve sulfur atoms per repeating unit. Such sulfur loading is consistent with the presence of short oligosulfide linkages, rather than extended polysulfide chains, which is in line with the high thermal stability (vide infra) and dark solid appearance of the polymeric materials.

PXRD confirmed the amorphous character of both polymers and the absence of remaining crystalline sulfur species such as $S_8$ (Figure S17 and S18). ToF-SIMS analysis further demonstrated the high purity of the materials, revealing only trace amounts of bromine residues from the starting materials in **FTN-Pol-$S_x$3** and sulfur fragments up to $S_4$ for both polymers, consistent with short covalently bound sulfur linkages within the polymer framework (Figure S23–S28). XPS analysis further corroborates these findings (Figure S29, Table S2 and S3). No residues of sodium salts for both polymers and only weak bromine species concentrations for **FTN-Pol-$S_x$3** (< 0.5 wt%) were detected. The survey spectra confirm the purity and defined composition of the polymers, with only minor O, F, and Si signals attributed to surface contamination. High-resolution XPS spectra of the core levels provide further insights into the chemical environment of both polymers. In the S 2p region, the main doublet for **FTN-Pol-$S_x$3** and **FTN-Pol-$S_x$6** with S $2p_{3/2}$ at 163.4 eV, characteristic of disulfide-type sulfur (C–S–S–C),[62] confirms the covalent incorporation into the polymer framework. In addition, doublets of weak intensity around 162 eV and 167 eV in both samples are attributed to terminal sulfur of the chain[63] and oxidized sulfur species ($SO_x$), respectively, likely formed by surface oxidation. The ratio between the central and terminal sulfur is two times higher for **FTN-Pol-$S_x$6** than for **FTN-Pol-$S_x$3**. This trend is consistent with the higher functionality of the **FTN-Pol-$S_x$6** monomer **1b** (six polymerizable sites vs three for **1a**). The latter leads to a bigger network with a smaller amount of terminal sulfur. In both cases the detection of both doublets indicates a quite limited length of the sulfur chain like already suggested by ToF-SIMS and EA. The N 1s spectra display a single peak at 400.5 eV for both **FTN-Pol-$S_x$3** and **FTN-Pol-$S_x$6**, comparable to the measured N 1s signal of the reference molecule **FTN-Pol**, indicating that the molecular building block remains chemically intact after polymerization (Table S2 and S3). The overall surface composition agrees well with the expected C/N/S ratios and supports the formation of a chemically uniform network. FT-IR spectra of **FTN-Pol-$S_x$3** and **FTN-Pol-$S_x$6** display distinct bands corresponding to aromatic (1000–1800 $cm^{-1}$) and aliphatic (2868–3061 $cm^{-1}$) C–H stretching vibrations, similar to the previously reported polymer **FTN-Pol**, confirming the preservation of the spiro-bridged *N*-HTA-based polymer backbone (Figure 3B).[56] Characteristic absorption bands in the 700–500 $cm^{-1}$ region are assigned to overlapping C–S and S–S stretching vibrations.[64] Both materials were also studied by Raman spectroscopy, revealing peaks at 485 $cm^{-1}$ for **FTN-Pol-$S_x$3** and 478 $cm^{-1}$ for **FTN-Pol-$S_x$6** that correlate with S–S stretching vibrations, in line with literature known disulfide compounds (Figure S11).[65,66] **FTN-Pol-$S_x$3** and **FTN-Pol-$S_x$6** were investigated by $^{13}$C MAS NMR (Figure S5 and S6). The signal at ~57 ppm is indicative for the C($sp^3$)-hybridized spiro-carbon and can be found in both compounds, again confirming the intact FTN backbone.[57,58]

TGA revealed a high thermal stability of both polymers, with only a minor initial weight loss below 150 °C (3–5%) attributed to moisture or residual solvent (Figure S12 and S13). The absence of a distinct decomposition step below 200 °C indicates that no free crystalline or long-chain polysulfur species are present, supporting the covalent incorporation of short sulfur linkages within the polymer framework. A gradual mass loss around 300 °C is attributed to the thermal cleavage of the *tert*-butyl substituents, followed by cleavage of C–S and S–S bonds within the sulfur chains between 350–600 °C. At higher temperatures (>600 °C), the polymer undergoes progressive decomposition and carbonization of the π-conjugated backbone fragments (Figure S14 and S15).

SEM micrographs of **FTN-Pol-$S_x$3** and **FTN-Pol-$S_x$6** show amorphous, coralloid aggregates composed of irregular fragments with rough, fractured surfaces and interparticle voids on the sub-µm scale (Figure 3A, Figure S20 and S21). No long-range order was observed, which is in line with PXRD results (vide supra). The morphology of both polymers was further studied by nitrogen sorption analysis at 77 K after activation at 100 °C for 2 h (Figure S30 and S31). In both cases, the nitrogen adsorption resulted in type II isotherms, showing little to no intrinsic porosity. In addition, no pronounced hysteresis between the adsorption and desorption isotherms was observed, which supports the presence of essentially non-porous, densely packed materials. The BET surface areas were determined to be 13.6 $m^2g^{-1}$ for **FTN-Pol-$S_x$3** and 46.3 $m^2g^{-1}$ for **FTN-Pol-$S_x$6**. These values are notably lower than that of the sulfur-free polymer **FTN-Pol** (690 $m^2g^{-1}$)[58], suggesting that the introduction of sulfur-based linkers leads to denser network packing and partial pore collapse during solvent removal, likely caused by enhanced interchain interactions, the high polarizability of sulfur, and the conformational flexibility of the oligosulfide chains.[67] Consequently, the accessible surface area is largely limited to interparticle voids rather than an intrinsic microporous framework.

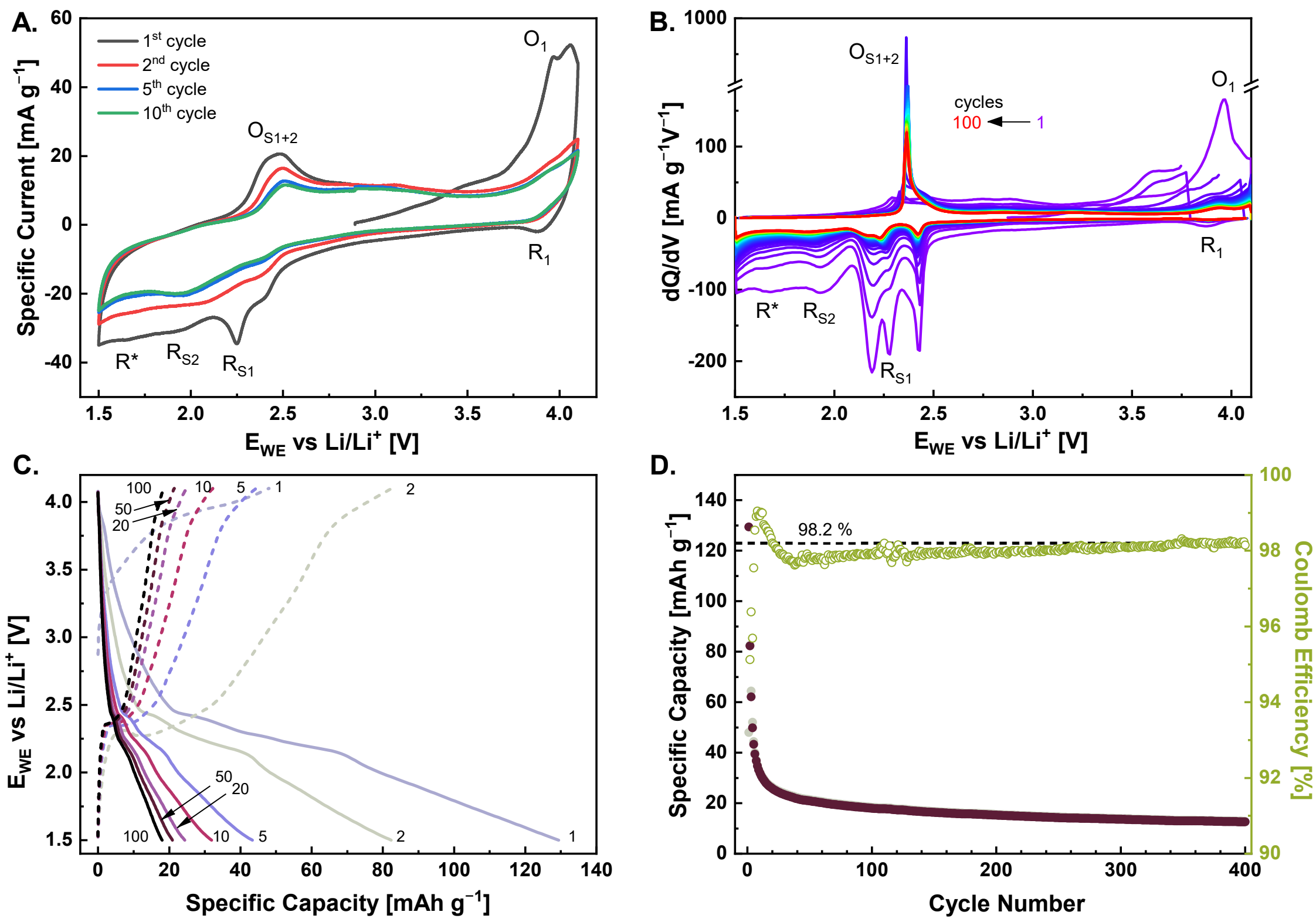

**Figure 4:** Electrochemical characterization of **FTN-Pol-S$_x$3** composite electrode (**FTN-Pol-S$_x$3**:CB:PVDF = 50:40:10 wt%) in 1 M LiTFSI solved in DOL:DME (1:1 v/v). A) CV at 0.1 mV s$^{-1}$ between 1.5 − 4.1 V vs. Li/Li$^+$ for 10 cycles. B) Differential capacity (dQ/dV) from galvanostatic cycling at 100 mA g$^{-1}$. C) Galvanostatic charge-discharge profiles at 100 mA g$^{-1}$. D) Cycling performance and Coulombic efficiency over 400 cycles. CB = carbon black; PVDF = poly(vinylidenefluoride); LiTFSI = lithium bis(trifluoromethanesulfonyl)imide, DOL = 1,3-dioxolane; DME = 1,2-dimethoxyethane. $O_{S1+2}$, and $R_{S1}$, $R_{S2}$ denote sulfur-related oxidation and reduction features, respectively.

The electrochemical properties of both sulfur-rich polymers were investigated with the aim to evaluate their potential for application as organic cathode materials. Initially, **FTN-Pol-S$_x$3** composite electrode **FTN-Pol-S$_x$3**:CB:PVDF (50:40:10 wt% was studied in a lithium half-cell configuration with 1 M LiTFSI in DOL:DME (1:1, v/v). Figure 4A shows the cyclic voltammetry response recorded at 0.1 mV s$^{-1}$ in the potential range of +1.5 to +4.1 V for 10 cycles. The open-circuit voltage (OCV) measures +2.89 V which is similar to our previously reported **FTN-Pol**.[58] During the first anodic sweep, there is a peak at +4.07 V ($O_1$) with a shoulder at +3.97 V indicating the oxidation of the FTN backbone. Compared to **FTN-Pol** ($O_1$ = +3.85 V), this feature is shifted to higher voltages and broadened, which suggests kinetic limitations possibly associated with the electronically insulating sulfur functionalities that may disrupt charge transport pathways. The corresponding reduction peak ($R_1$) at +3.88 V is strongly suppressed, indicating pronounced redox asymmetry. This behavior may arise from sluggish charge-transfer kinetics, partial electrode passivation, or competing parasitic reactions. In general, the FTN redox process in **FTN-Pol-$S_x$3** is centered around +3.97 V and exhibits an overpotential of 0.2 V which is nearly three times larger than the overpotential observed in sulfur-free **FTN-Pol**, thus supporting the assumption of kinetic limitations due to the incorporated sulfur. In addition to these FTN-related signals, two reduction peaks at +2.25 V ($R_{S1}$) and +1.95 V ($R_{S2}$) are observed. In line with our reference measurements (Figure S33) and literature reports on polysulfides,[68] these signals can be assigned to sulfur redox activity. Both features diminish significantly within the first five cycles but remain detectable in the 10th cycle. A further low-potential feature at ≈ +1.5 V (R*) cannot be assigned to either FTN- or sulfur-related processes. On the anodic sweep, a broad oxidation peak at +2.49 V ($O_{S1+2}$) appears, which follows a similar trend of rapid initial decay with subsequent stabilization. The overall poor reversibility of $R_{S1}$, $R_{S2}$, and R* suggests substantial sulfur dissolution and possible polysulfide shuttle phenomena, consistent with previous studies of Li-S systems.[20,68] While widely used, it should be noted that under strongly oxidizing conditions, ether-based electrolytes such as 1,3-dioxolane may undergo side reactions, which could contribute to irreversible processes, particular when extending the voltage window above 4 V. Although no direct evidence for significant electrolyte decomposition was observed in the present study, its contribution cannot be entirely excluded and should be considered when interpreting the electrochemical data. To further investigate the redox mechanisms, the differential capacity derived from galvanostatic cycling at 100 mA $g^{-1}$ is presented in Figure 4B. The OCV of +2.87 V is consistent with the CV measurements. A distinct oxidation peak at +3.96 V ($O_1$) and a broad, weak reduction peak at +3.87 V ($R_1$) confirm the redox process located at the FTN backbone. $R_1$ decays rapidly within the first five cycles but persists at low intensity up to 400 cycles, yielding an average redox potential ($E_{1/2}$) of +3.92 V with a reduced overpotential of 0.09 V. In the lower-potential region, multiple overlapping reduction peaks at +2.43, +2.28, and +2.19 V (collectively denoted $R_{S1}$) are identified and are attributed to sulfur conversion reactions. These signals are initially pronounced but decline within the first five cycles, stabilizing at low intensity thereafter. A broad reduction feature near +1.94 V (denoted $R_{S2}$) extends toward +1.5 V (R*) and suggests deeper sulfur reduction. The corresponding oxidation $O_{S1+2}$ appears as a broad feature centered around +2.4 V in the first cycle, undergoes sharpening in subsequent cycles and as such reflects partially reversible sulfur electrochemistry. The potential profiles shown in Figure 4C provide further insights into the individual redox contributions. FTN oxidation ($O_1$) delivers ≈ 50 mA h $g^{-1}$, while its reduction ($R_1$) accounts for only ≈ 4 mA h $g^{-1}$. The sulfur redox processes ($R_{S1}$ and $R_{S2}$/R*) contribute ≈ 100 mA h $g^{-1}$, resulting in an initial discharge capacity of 129 mA h $g^{-1}$. The oxidation peak $O_{S1}$ recovers ≈ 33 mA h $g^{-1}$, indicating only partial reversibility. All redox features exhibit rapid capacity fading, retaining ≈ 25% of the initial capacity after 10 cycles and ≈ 10% after 400 cycles, consistent with continuous sulfur loss and shuttle effects in the ether-based electrolyte. Long-term cycling performance further supports these findings. The initial discharge capacity of 129 mA h $g^{-1}$ (0.13–0.19 mAh $cm^{-2}$) decays exponentially to 31 mA h $g^{-1}$ (0.03–0.05 mAh $cm^{-2}$) after 10 cycles and to 16 mA h $g^{-1}$ (0.02–0.03 mAh $cm^{-2}$) after 200 cycles, reaching 13 mA h $g^{-1}$ (0.01–0.02 mAh $cm^{-2}$) at cycle 400. The Coulombic efficiency increases from 37.2% in the first cycle to 99.0% by the 10th cycle, decreases slightly to 97.6% at cycle 40, and subsequently converges to 98.2% at longer cycling. The absence of capacity stabilization highlights severe polysulfide shuttle effects and overall instability of **FTN-Pol-S$_x$3** composite system (for alternative electrolytes tested under the same conditions, see Figure S35).

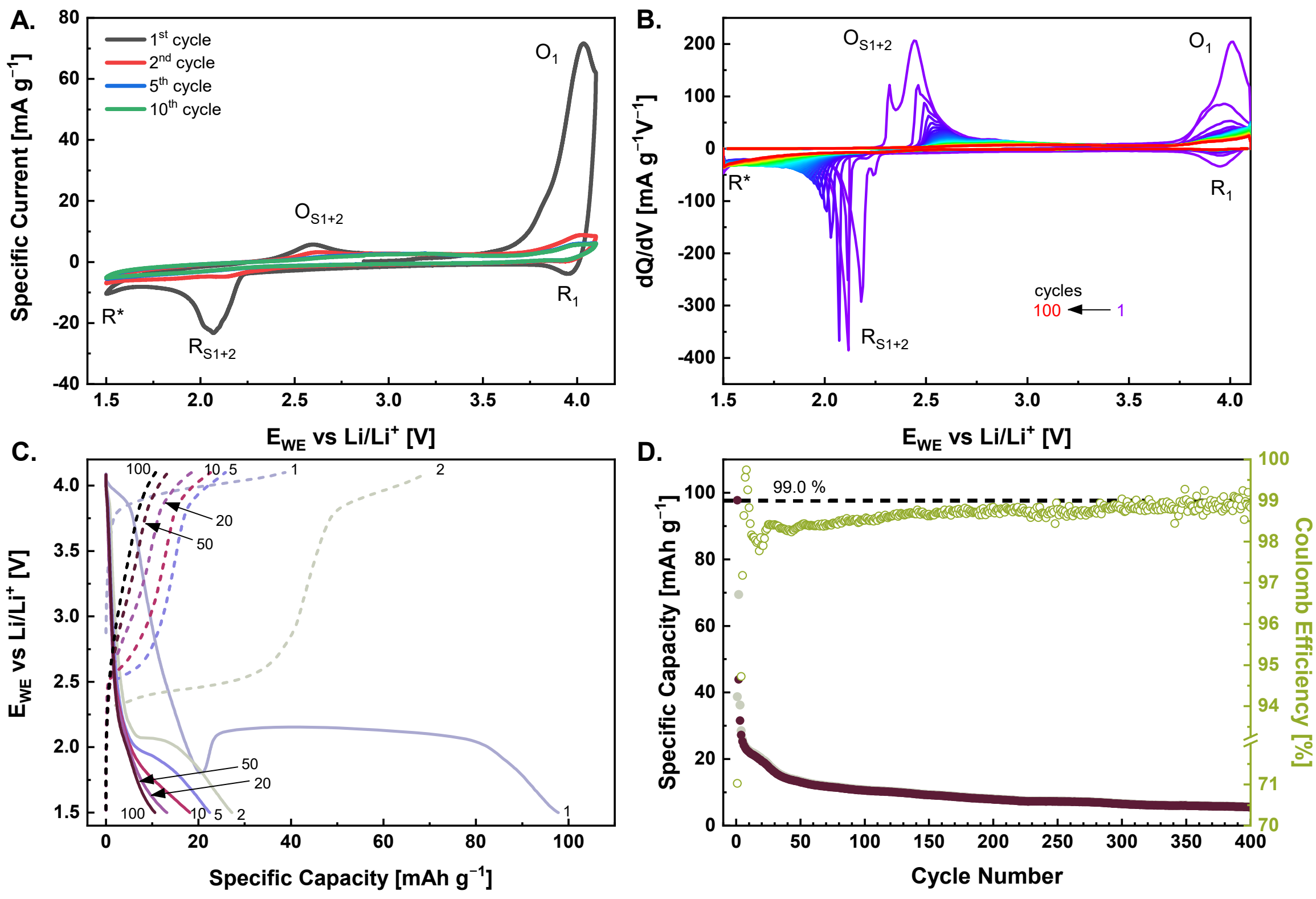

**Figure 5:** Electrochemical characterization of **FTN-Pol-S$_x$6** composite electrode (**FTN-Pol-S$_x$6**:CB:PVDF = 50:40:10 wt%) in 1 M LiTFSI solved in DOL:DME (1:1 v/v). A) CV at 0.1 mV $s^{-1}$ between 1.5 − 4.1 V vs. $Li/Li^+$ for 10 cycles. B) Differential capacity (dQ/dV) from galvanostatic cycling at 100 mA $g^{-1}$. C) Galvanostatic charge-discharge profiles at 100 mA $g^{-1}$. D) Cycling performance and Coulombic efficiency over 400 cycles. CB = carbon black; PVDF = poly(vinylidenefluoride); LiTFSI = lithium bis(trifluoromethanesulfonyl)imide, DOL = 1,3-dioxolane; DME = 1,2-dimethoxyethane. $O_{S1+2}$, and $R_{S1+2}$ denote sulfur-related oxidation and reduction features, respectively.

Electrochemical measurements on the composite electrode with **FTN-Pol-S$_x$6** show similar electrochemical behavior (Figure 5 and Table 1). Specifically, the data demonstrate similar electrochemical processes associated with the FTN backbone structure as discussed above for **FTN-Pol-$S_x$3**, but the corresponding redox process exhibits a lower overpotential of 0.1 V. In the sulfur region, the two distinct reduction features observed for **FTN-Pol-$S_x$3** merge into a single process at +2.07 V ($R_{S1+2}$). Moreover, the differential capacity data shown in Figure 5B imply an additional irreversible shoulder-like feature at +2.24 V which decays within the first 60 cycles and gradually shifts to lower voltages before stabilizing at ≈ 10 mA $g^{-1}V^{-1}$ up to 400 cycles. On oxidation, $O_{S1+2}$ appears as a broad peak centered around +2.44 V with a shoulder at +2.32 V, which shifts to higher potentials upon continued cycling. This does not change the redox potential ($E_{1/2}$) but leads to a pronounced increase in overpotential from 0.27 V in the 1st cycle to 0.90 V by the 20th cycle. This evolution suggests progressively hindered kinetics, likely associated with sulfur dissolution into the electrolyte. The galvanostatic charge-discharge profiles and the long-term cycling data in Figure 5C and D underline the similar behavior to **FTN-Pol-$S_x$3**. Notably, the Coulombic efficiency spread increases substantially beyond 300 cycles, reflecting growing instability of the redox processes. Taken together, these results show that the **FTN-Pol-$S_x$6** composite suffers even more from unstable sulfur redox activity and pronounced long-term degradation than **FTN-Pol-$S_x$3**.

Although the synthesis of both polymers is simple and high yielding, which makes them in principle interesting for practical applications, their electrochemical performance is rather disappointing, most likely due to their poorly defined structure and the redox instability of the oligosulfide linkers. For this reason, we set out to synthesize a well-defined analogue in which the sulfur atoms are embedded in π-conjugated terthiophene moieties to provide for stabilization. In this context, terthiophene was selected as a structurally defined, π-conjugated sulfur-containing linker that represents an extended analogue of the flexible oligosulfide chains used in the previous polymers. This design enables a direct comparison between flexible polysulfide linkers and π-conjugated thiophene-based units within the same FTN framework.

### *2.3. Synthesis and Characterization of the Terthiophene-Linked Polymer **FTN-Pol-TTP***

To maximize the sulfur content in the resulting material, the π-conjugated polymer **FTN-Pol-TTP** was synthesized exclusively from the hexabrominated monomer **1b** via Stille cross-coupling polymerization with a stannylated terthiophene linker (Figure 6). The required bis(trimethylstannyl)terthiophene was obtained in quantitative yield from commercially available terthiophene by successive lithiation and stannylation using $^{n}$BuLi and trimethyltin chloride at 0 °C. Subsequent polymerization of **1b** with the stannylated linker in toluene using [Pd($PPh_3$)$_4$] as catalyst, followed by Soxhlet extraction (methanol, THF, toluene, and $CH_2Cl_2$) and drying under reduced pressure, afforded **FTN-Pol-TTP** as an orange powder in 88% yield. The polymer is insoluble in water and common organic solvents.

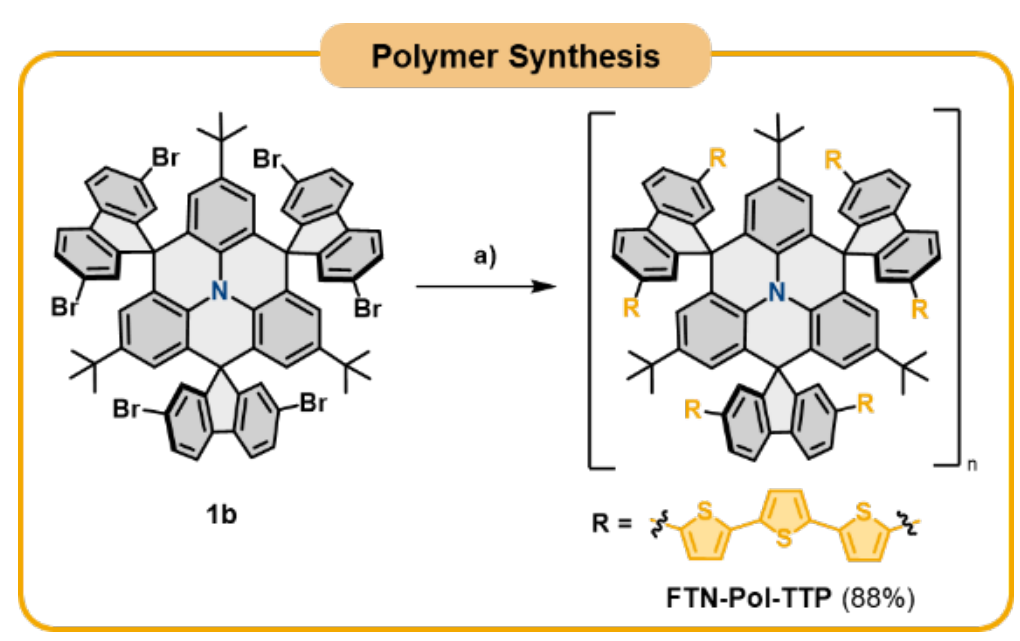

**Figure 6:** Synthetic route toward terthiophene-linked polymer **FTN-Pol-TTP**. Reagents and conditions: a) $[1^2,2^2:2^5,3^2$-terthiophene]-$1^5,3^5$-diyl)bis(trimethylstannane), $[Pd(PPh_3)_4]$, toluene, 130 °C, 60 h, $N_2$. For the synthesis of $[1^2,2^2:2^5,3^2$-terthiophene]-$1^5,3^5$-diyl)bis(trimethylstannane), see the Supporting Information.

As shown in Figure 7A, photographs of the purified polymer demonstrate its amorphous solid, orange appearance. The same characterization methods as for **FTN-Pol-S$_x$3** and **FTN-Pol-S$_x$6** were employed to verify the purity, composition, structure, and morphology of **FTN-Pol-TTP**. Elemental combustion analysis is in good agreement with the calculated composition of the defined terthiophene-linked framework, confirming the successful polymerization of the brominated precursor **1b** (calculated: S, 13.87% vs. found: S, 11.34%). ToF-SIMS and PXRD demonstrate the high purity and amorphous nature of the material, with no remaining palladium species and only trace signals of tin and bromine originating from incomplete conversion of the starting materials (Figure S19, S27, and S28). In addition, XPS analysis confirms the purity and composition of **FTN-Pol-TTP**, showing major signals for C–C/C–H (285.0 eV), thiophene-based sulfur (164.4 eV), and N 1s (400.6 eV), with only minor surface contamination (Figure S29 and Table S4). Comparison of the FT-IR spectra of **FTN-Pol-TTP** with the reference polymer **FTN-Pol** and the monomeric reference **TTP** reveals the expected aromatic (1800–800 $cm^{-1}$) and aliphatic (3100–2600 $cm^{-1}$) vibrations and distinct C–S stretching vibrations between 750–500 $cm^{-1}$, consistent with the **TTP** model compound (Figure 7A). Furthermore, Raman spectroscopy shows a characteristic band at 1454 $cm^{-1}$, which corresponds to the strong IR band at 1455 $cm^{-1}$ and is assigned to C=C stretching within the extended π-systems (Figure S11).[69] According to TGA, **FTN-Pol-TTP** shows high thermal stability up to 350 °C (Figure S14). An initial minor weight loss below 150 °C is attributed to moisture and gases, followed by a gradual mass decrease between 250 °C and 350 °C corresponds to the thermal elimination of the *tert*-butyl substituents. Major decomposition between 350 °C and 600 °C assigned to cleavage of C–S and C–C bonds within the terthiophene linkers. First, above 600–700 °C, degradation and carbonization of the conjugated spiro-FTN backbone take place. SEM micrographs reveal amorphous, irregular aggregates composed of irregular fragments with rough surfaces and no apparent long-range order (Figure 7A and Figure S22). Nitrogen sorption measurements at 77 K show type II isotherms with a BET surface area of 45 $m^2g^{-1}$, indicating low intrinsic porosity and dense packing of the polymer network (Figure S32).

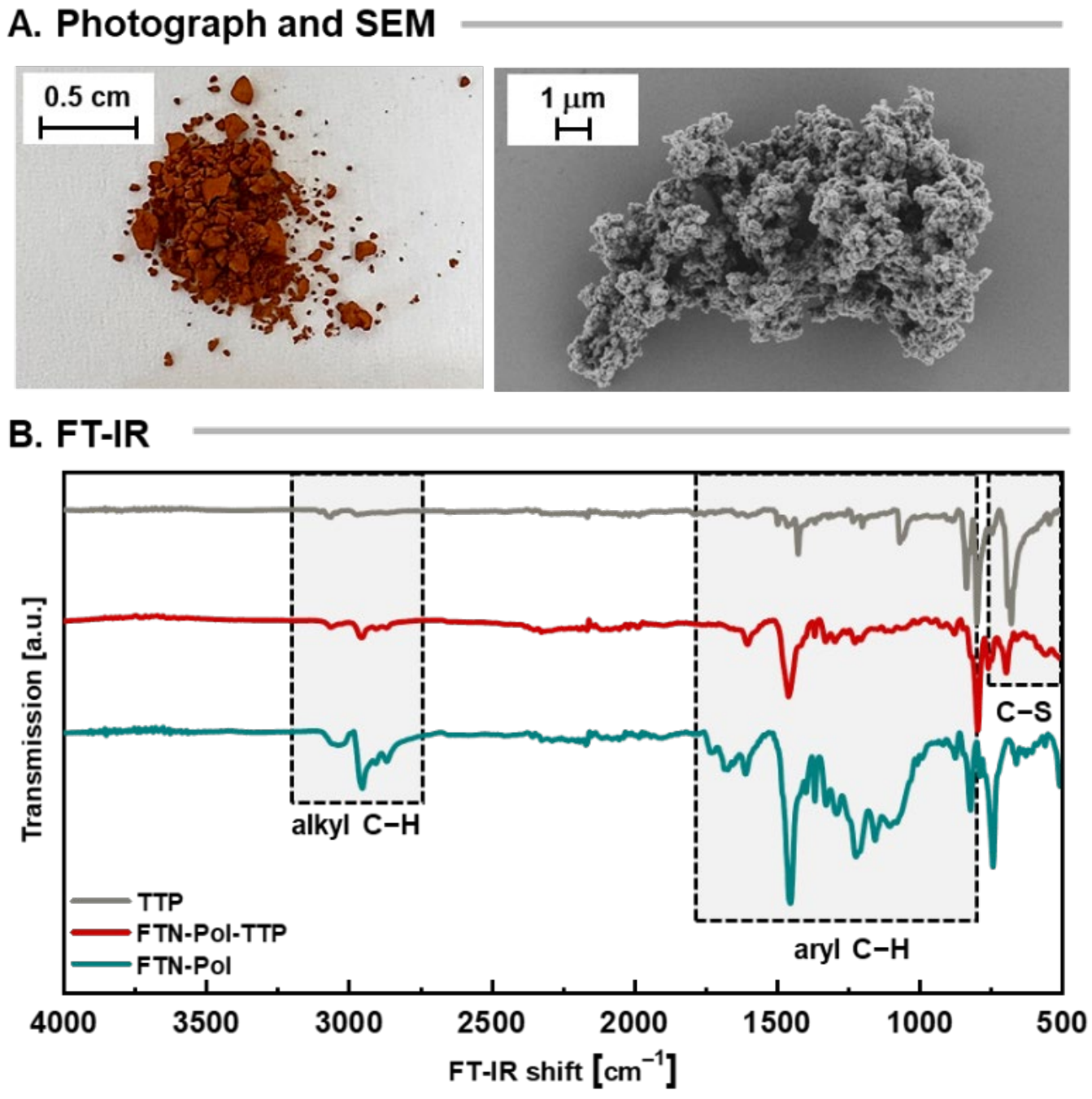

**Figure 7:** Morphological and structural characterization of polymer **FTN-Pol-TTP**. (A) Photograph (left) and SEM image (right) of the powder; (B) FT-IR spectra of the polymer and the model compounds **FTN-Pol** and **TTP**.

To evaluate the impact of terthiophene linkers on the electrochemical properties, **FTN-Pol-TTP** was characterized in composite electrodes **FTN-Pol-TTP**:CB:PVDF (50:40:10 wt%). Measurements were conducted in a carbonate-based electrolyte ($LiPF_6$ in EC/DMC) as **FTN-Pol-TTP** behaves as a π-conjugated organic redox polymer rather than a sulfur conversion-type material. This choice reflects that the electrochemical response is not primarily based on classical polysulfide-mediated conversion processes. **FTN-Pol-TTP** exhibits electrochemical characteristics comparable to the **FTN-Pol-S$_x$3** electrodes (Table 1), yet predominantly transient **TTP**-centered redox processes are observed, while the FTN backbone maintains its intrinsic electrochemical stability and reversibility (Figure 8). As shown in Figure 8A and B, an additional oxidation feature (O*) emerges from the 2nd cycle below +3.46 V without a corresponding reduction peak, suggesting an irreversible oxidation process associated with the TTP moieties, consistent with literature reports on oligothiophene oxidation.[89] The reversible high-voltage feature at +3.86 V is attributed to the FTN backbone.

Minor low-potential redox couples ($O_{S1}/R_{S1}$ and $O_{S2}/R_{S2}$) at approximately +2.4 V and +2.7 V, respectively, are assigned to sulfur-related processes. No clear reduction feature attributable to terthiophene is observed, in agreement with literature reports indicating limited reversible reduction of terthiophene units.[90,91] This O* feature shifts to higher potentials up to +3.63 V and increases in intensity before disappearing, which is consistent with its transient nature. In the initial cycles, the oxidation process $O_1$ in the high-voltage region shows irreversibility, which coincides with the presence of the O* feature. Upon its disappearance, $O_1$ becomes more reversible, indicating that the remaining electrochemical activity in this voltage region is primarily dominated by the FTN backbone. Minor low-potential redox couples ($O_{S1}/R_{S1}$ and $O_{S2}/R_{S2}$) vanish within the first five cycles, and a pronounced reduction peak near +1.5 V (R*) is observed, analogous to **FTN-Pol-S$_x$3** but more intense, as confirmed by the galvanostatic charge-discharge curves (Figure 8C). The long-term cycling behavior reveals an initial capacity of 101 mA h $g^{-1}$ (0.10–0.15 mAh $cm^{-2}$) comparable to the other FTN-based polymers (see Table 1), followed by the characteristic rapid capacity decay. Despite this decay, the Coulombic efficiency approaches 99.7% and remains stable throughout extended cycling (Figure 8D). Overall, the TTP moieties appear to contribute only short-lived, degradative redox activity, whereas the FTN backbone dominates the sustained reversible performance, consistent with the behavior reported previously for sulfur-free **FTN-Pol**.[58]

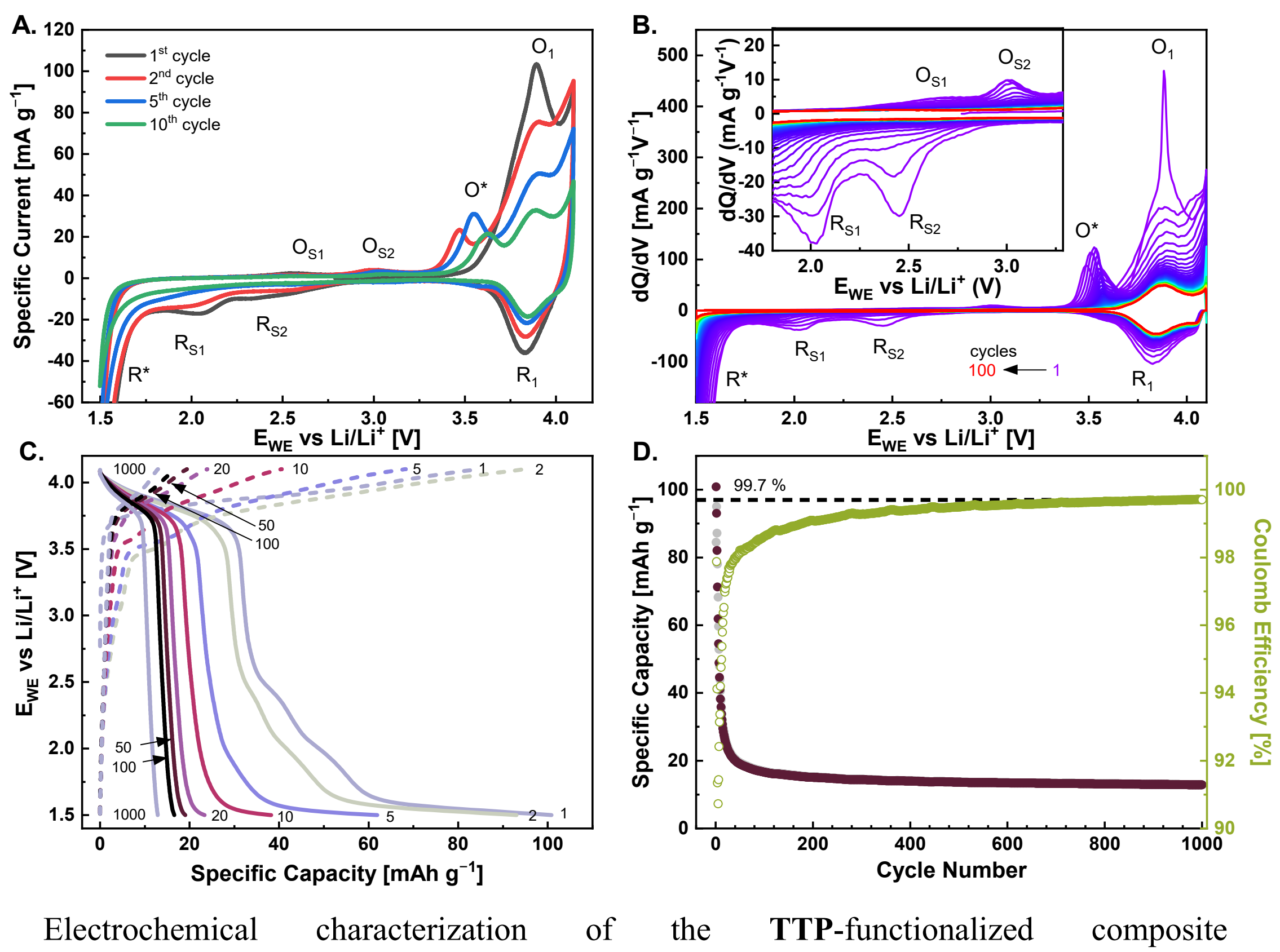

**Figure 8:** Electrochemical characterization of the **TTP**-functionalized composite electrode

(**FTN-Pol-TTP**:CB:PVDF = 50:40:10 wt%) in 1 M $LiPF_6$ solved in EC:DMC (1:1 v/v). A) CV at 0.1 mV $s^{-1}$ between 1.5 − 4.1 V vs. $Li/Li^+$ for 10 cycles. B) Differential capacity plots (dQ/dV) from galvanostatic cycling at 100 mA $g^{-1}$. C) Galvanostatic charge-discharge profiles at 100 mA $g^{-1}$. D) Cycling performance and Coulombic efficiency over 1000 cycles. $O_{S1}$, $O_{S2}$ and $R_{S1}$, $R_{S2}$ denote sulfur-related oxidation and reduction features, respectively.

**Table 1** Comparative summary of electrochemical performance of previously reported **FTN-Pol**[58] and sulfur-rich polymers FTN-based polymers as cathode materials in Li-ion half-cells obtained from cyclic voltammetry (CV) and galvanostatic cycling with potential limitation (GCPL) studies presented in Figures 4, 5, and 8.

| | $C_{theo}$[2] [mAh $g^{-1}$] | $C_{ini}$[3] [mAh $g^{-1}$] | $C_{@10}$[4] [mAh $g^{-1}$] | $C_{@400}$[4] [mAh $g^{-1}$] | *CE* [%] | $E_{1/2}$ [V] | *OvPot*[5] [V] |
|---|---|---|---|---|---|---|---|
| **FTN-Pol**[58] | 30 | 26 | 22 | 21 | 99.7 | 3.81 | 0.02 |
| **FTN-Pol-S$_x$3** | 440 | 129 | 31 | 13 | 98.2 | 3.92 | 0.09 |
| **FTN-Pol-S$_x$6** | 424 | 98 | 22 | 8 | 99.0 | 3.98 | 0.03 |
| **FTN-Pol-TTP** | 48 | 101 | 38 | 14 | 99.7 | 3.86 | 0.04 |

A comparative evaluation of the sulfur-rich polymers **FTN-Pol-S$_x$3**, **FTN-Pol-S$_x$6**, and **FTN-Pol-TTP** against the unmodified **FTN-Pol** highlights clear differences in theoretical capacity, practical utilization, and cycling stability (Table 1 and Figure S36). In general, sulfur functionalization in **FTN-Pol-S$_x$3** and **FTN-Pol-S$_x$6** substantially increases the theoretical capacity relative to **FTN-Pol**, whereas terthiophene moieties in **FTN-Pol-TTP** provide only a moderate enhancement. The initial capacities follow the same trend, with **FTN-Pol-S$_x$3** delivering the highest value (129 mA h $g^{-1}$), followed by **FTN-Pol-TTP** (101 mA h $g^{-1}$), and **FTN-Pol-S$_x$6** (98 mA h $g^{-1}$), all clearly exceeding the performance of sulfur-free polymer **FTN-Pol** (26 mA h $g^{-1}$) by far. These values correspond to initial utilizations of approximately 23–29% for **FTN-Pol-S$_x$3** and **FTN-Pol-S$_x$6** with respect to their theoretical capacities. For **FTN-Pol-TTP**, the initial capacity exceeds the theoretical capacity estimated for the terthiophene-based one-electron redox process (48 mAh $g^{-1}$). As all three materials exhibit substantial capacity fading upon cycling, these values represent initial rather than sustained utilizations. Possible contributions from irreversible side reactions, or interfacial processes may account for part of the excess initial capacity observed for **FTN-Pol-TTP**. Capacity retention at 10th cycle indicates that **FTN-Pol-S$_x$3** surpasses not only both sulfur-rich **FTN-Pol-S$_x$6** and **FTN-Pol-TTP** but also pristine **FTN-Pol**. At extended cycling up to 400 cycles, all materials converge to comparatively low capacities, demonstrating the pronounced long-term fading characteristic of the sulfur-functionalized systems. This low utilization and rapid decay may be attributed to a combination of limited electronic conductivity, restricted ion transport within the polymer matrix, and incomplete accessibility or irreversible loss of sulfur based redox-active sites due to partial dissolution of sulfur species into the electrolyte and possible shuttle effects. Coulombic efficiencies, however, remain consistently high (98.0–99.7%), reflecting stable charge-discharge reversibility despite capacity decay. The FTN-related redox potentials increase upon sulfur and TTP incorporation, with **FTN-Pol-$S_x$6** showing the highest value (+3.98 V), followed by **FTN-Pol-$S_x$3** (+3.92 V), and **FTN-Pol-TTP** (+3.86 V), compared to +3.81 V for the parent polymer **FTN-Pol**.[58] Overpotentials remain low, with **FTN-Pol** and **FTN-Pol-$S_x$6** exhibiting the most favorable kinetics (0.02 and 0.03 V), while **FTN-Pol-$S_x$3** displays the highest (0.09 V). Overall, sulfur and TTP incorporation significantly boosts theoretical and initial capacities but accelerates long-term degradation. These results indicate that neither sulfur nor TTP moieties provide sustainable electrochemical benefit and that the long-term performance is predominantly governed by the excellent intrinsic stability of the FTN backbone.

[2] Theoretical specific capacity, calculated as described in the Supporting Information
[3] Initial specific capacity (first cycle)
[4] Specific capacity at the indicated cycle
[5] Overpotential

For a broader contextualization, the electrochemical performance of the present FTN-based polymers was compared with representative covalently bound sulfur-containing polymer cathodes reported in the literature (Table S5).[70–72] Compared to these literature systems, the FTN-based polymers show significantly lower reversible capacities and more pronounced capacity fading, while maintaining consistently high Coulombic efficiencies, highlighting that the electrochemical response is predominantly governed by the stable FTN redox backbone rather than by sustained sulfur-based conversion processes.

At the same time, the observed rapid capacity fading and transient linker-based redox processes highlight important limitations in the current systems and underline the complexity of interpreting the electrochemical behavior of sulfur-containing polymer frameworks. Future studies may therefore focus on stabilizing the linker-based redox motifs, for example by improving the confinement of sulfur species within the polymer network,[73] modifying the electronic structure of the linker units,[74] or optimizing the electrode formulation and electrolyte composition for energy-storage applications.[75]

## 3. Summary

We have synthesized sulfur- and terthiophene-functionalized spirofluorene-bridged *N*-heterotriangulene (FTN) polymers through $S_NAr$ and Stille cross-coupling polymerization, enabling the incorporation of short oligosulfide linkages or well-defined terthiophene units into the FTN scaffold. Comprehensive structural characterization by elemental analysis, ToF-SIMS, FT-IR and Raman spectroscopy, XPS, and solid-state NMR confirms the efficient incorporation of the sulfur-based substituents while preserving the spiro-bridged FTN backbone. All polymers are amorphous, insoluble in common organic solvents, and thermally robust solids with sulfur contents of up to 25 wt%. Electrochemical studies in lithium-ion half-cells reveal that sulfur linking markedly increases theoretical and initial capacities. However, the conversion-type redox chemistry results in rapid capacity fading and limited reversibility, consistent with polysulfide dissolution in the electrolyte. In contrast, terthiophene substitution appears to be associated with predominantly transient redox processes, while the intrinsic high-voltage redox-activity of the FTN unit remains comparatively stable and reversible with efficiencies approaching 99.7%. Our findings demonstrate that sulfur and terthiophene linking motifs can boost initial charge-storage capacity but do not provide sustainable electrochemical benefit, as the long-term cycling is governed predominantly by the intrinsic redox stability of the FTN scaffold. These observations provide insight into how different sulfur-containing linkers influence the redox behavior of FTN-based polymer frameworks and highlight important structure-property relationships within this class of redox-active materials. Moreover, we anticipate that polymerization of FTN using functional redox-active linkers will open avenues towards robust organic cathode materials with enhanced electrochemical properties.

**Supporting Information**

The authors have cited additional references within the Supporting Information.[76–88]

**Author contributions**

A.J. synthesized and characterized all building blocks and polymers, contributed to the data analysis, and wrote the original manuscript draft. T.W. took part in the conceptualization of the project, designed and performed the electrochemical studies, analyzed these data, and wrote together with R.K. the original draft of the respective manuscript part. S.L. did the Raman measurements, analyzed these data together with J.Z., and plotted the spectra. A.W. did the ToF-SIMS measurements and analyzed these data. V.T. analyzed the polymers via XPS and analyzed these data and provided the spectra. I.W. and R.C. characterized the polymers by SEM, and together with R.R.S. analyzed the data and prepared the images. M. K. conceived and supervised the project, acquired the funding, edited, and finalized the manuscript. All authors discussed the results and commented on the manuscript.

## Acknowledgments

The generous funding by the Deutsche Forschungsgemeinschaft (DFG) – Project number 281029004 – SFB 1249 is acknowledged. T.W. and R.K. acknowledge the support by the DFG via the Research Training Group GRK 2948/1 'Mixed Ionic Electronic Transport'. The authors thank Dr. Frank Hampel from the Friedrich-Alexander-Universität Erlangen-Nürnberg for measuring elemental analysis. The authors thank Dr. Jürgen Gross from Heidelberg University for measuring $^{13}C$ CP MAS NMR spectra.